\documentclass[a4paper]{EslabStyle}
\usepackage{graphics}

\def\msun{M_{\sun}}
\def\lsun{L_{\sun}}
\def\kms{\rm \, km \, s^{-1}}
\newbox\grsign \setbox\grsign=\hbox{$>$} \newdimen\grdimen \grdimen=\ht\grsign
\newbox\simlessbox \newbox\simgreatbox
\setbox\simgreatbox=\hbox{\raise.5ex\hbox{$>$}\llap
     {\lower.5ex\hbox{$\sim$}}}\ht1=\grdimen\dp1=0pt
\setbox\simlessbox=\hbox{\raise.5ex\hbox{$<$}\llap
     {\lower.5ex\hbox{$\sim$}}}\ht2=\grdimen\dp2=0pt
\def\gtrsim{\mathrel{\copy\simgreatbox}}
\def\lessim{\mathrel{\copy\simlessbox}}

\begin{document}

\title{STAR FORMATION: LESSONS FROM TAURUS}

\author{Lee Hartmann}
\institute{Harvard-Smithsonian Center for Astrophysics, 60 Garden St., MS 42,
Cambridge, MA 02138}

\maketitle

\begin{abstract}
The Taurus molecular cloud complex is the paradigm for quiescent, low-density,
isolated star formation.  Yet the age distribution of its stellar population
indicates that star formation is a rapid and dynamic process, inconsistent with
the old picture of magnetically-controlled protostellar cloud collapse. Instead,
Taurus seems to have formed stars in a manner qualitatively consistent 
with the rapid cloud formation and dispersal inferred for other, higher-density 
star-forming regions.  I suggest that the Taurus clouds were formed rapidly by
the collision of atomic gas streams in the interstellar medium, and that star formation
ensued immediately afterward.
\keywords{molecular clouds, Stars: formation, Stars: X-rays}
\end{abstract}
{\footnotesize Proc.\ of 33rd ESLAB Symp.\ \emph{``Star
formation from the small to the large scale''} (F.~Favata,
A.\,A.~Kaas \& A.~Wilson eds., ESA SP-445, 2000)}

\section{Introduction}
The Taurus-Auriga molecular cloud complex has long been considered an important
region for testing our understanding of star formation.  Because of
its proximity, position in the northern celestial hemisphere, and low values of dust extinction,
Taurus has probably the best-characterized stellar population of all star-forming
regions, despite its large angular size on the sky.  Though the Taurus
clouds may not be typical of star forming regions in the Milky Way,
the relatively large spatial separations between young stars makes the region
particularly well-suited to the study of ``isolated'' star formation, and therefore
it presumably provides particularly simple and direct tests of theory.

In the paradigm of isolated, low-density, low-mass star formation that has
dominated the field for more than a decade (e.g., Shu, Adams, \& Lizano 1987),
molecular cloud ``cores'' cannot collapse rapidly to form stars because of the
restraining forces of magnetic fields.  The gas must first diffuse across magnetic
field lines, as permitted by the low ionization state of dark clouds, before enough
magnetic flux can be removed to allow collapse to proceed.   Depending upon how
``subcritical'' or magnetically-dominated the cloud core is, the precise ionization state,
etc., the timescale for this ambipolar diffusion process is often estimated to be
of order 5-10 Myr (Mouschovias 1991).  This timescale is much longer than the dynamical
timescale $\sim {\rm few} \times 10^5$~yr of a cloud core, and so the core can be thought of as in
approximate hydrostatic equilibrium.  Slowing the collapse of molecular clouds 
by magnetic forces was thought to help explain the low rate of galactic star formation 
compared to the maximum rate that would occur if all molecular clouds are collapsing
at free-fall rates (Zuckerman \& Evans 1974).

It has become increasingly apparent that this picture of slow star formation   
has severe problems in explaining the rapidity of star formation in dense
regions.  Indeed, Shu et al. (1987) had already
recognized this problem, and suggested that formation of clusters
proceeds supercritically, i.e. the magnetic field is not strong enough to prevent collapse, and
ambipolar diffusion is therefore not required.  Originally, Shu et al. suggested that
supercritical collapse was the mode of formation of high-mass stars, since these appear
preferentially in dense, populous clusters.  However, we now recognize that high-mass
stars essentially always appear with corresponding large numbers of low-mass stars
(as illustrated best by the Orion Nebula Cluster (ONC): Herbig \& Terndrup 1986; Hillenbrand 1997), 
so that to suppose that high-mass stars form supercritically is to concede that 
large numbers of low-mass stars form that way as well.

What seems to have been overlooked is that the Taurus molecular cloud complex {\em also}
shows every evidence for rapid triggering of 
star formation, despite being the paradigm region for slow collapse.
Indeed, in some ways Taurus actually poses a stiffer test for star formation theories than the ONC;
though the timescale of coordination may be a bit longer in Taurus
($\sim$ 1-2 Myr vs. $\sim 0.5$ Myr for the ONC), the distance scale over which star formation
is coordinated is much larger in Taurus ($\sim 20$ pc vs. $\sim 2$ pc).

\begin{figure*}[t]
\resizebox{18cm}{!}{\includegraphics{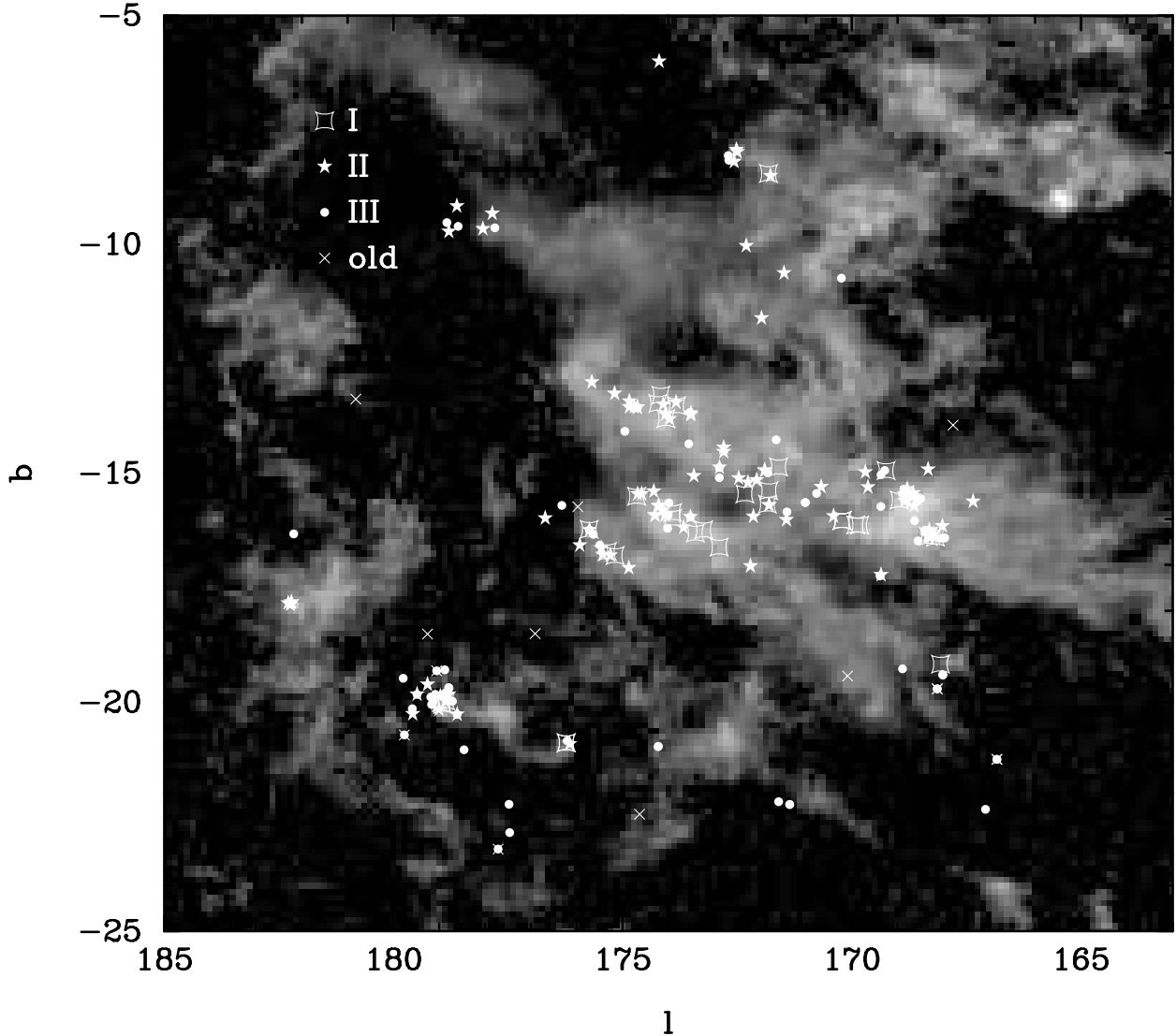}}
\caption{Young stars in the Taurus region superimposed on the
$^{12}$CO map of the Taurus region (grayscale) from Megeath, Dame, \& Thaddeus.
The stellar objects are divided into
Class I (protostars; heavily-extincted young stars),
Class II (accreting T Tauri stars), and Class III (non-accreting T Tauri stars).
``Old'' stars are objects from Walter et al. (1988) which are estimated to have
ages greater than 10 Myr, plus objects labeled ``post-T Tauri stars''
by Magazzu \& Martin (1998) (see \S 5).
}
\end{figure*}

In this contribution I review the age distribution of the young stars in Taurus
cloud complex.  I show that neither uncertainties in determining stellar ages nor the
widely-dispersed X-ray population in the general vicinity contradict the conclusion that
there has been a rapid triggering of star formation in Taurus.  The short timescales
involved indicate that ambipolar diffusion is unlikely to control the
rate of star formation.   
I suggest that the Taurus clouds have formed rapidly from converging
flows in the interstellar medium, a process which is probably applicable to high-density star-forming
regions as well.  The general picture is one in which 
molecular clouds both form and disperse rapidly; 
the limited rate of star formation
on galactic scales is the product of low efficiencies in converting gas into stars rather than
the result of slow collapse.

\section{Stellar Ages in Taurus}

Figure 1 shows the spatial distribution of young stars in the direction of Taurus,
superimposed on the new $^{12}$CO map of the region by Megeath, Dame, \& Thaddeus (personal
communication).  As previously noted, the positions of the young stars correlate very well 
with the presence of molecular gas.  The stars are clearly concentrated in filamentary structures,
with a few small groups included.  Most of the Taurus stars can be encompassed with a region
of diameter $\sim 20$~pc, although the entire span over which newly-formed stars can be
found approaches 40~pc.  

Figure 2 shows the distribution of ages among these stars (those with
sufficient information to be placed in the HR diagram).  For this purpose the stellar properties
were taken from the compilation of Kenyon \& Hartmann (1995), supplemented by additional objects
from Brice\~no et al. (1997, 1998, 1999).  The ages were determined from the 
``CMA'' evolutionary tracks
of D'Antona \& Mazzitelli (1994) assuming a distance to the complex of 140 pc
(Kenyon, Dobryzcka, \& Hartmann 1994).  With these data and assumptions, 
the median age of the Taurus pre-main sequence stars is about 1 Myr.
Note that there is little difference in the age distributions of the accreting ``classical''
T Tauri stars, or CTTS, which have strong H$\alpha$ emission and show infrared and ultraviolet
excess continuum emission, and the ``weak-emission'' T Tauri stars or WTTS, which are not
accreting from disks and show only weak H$\alpha$ emission and other signatures of enhanced
solar-type magnetic activity.

\begin{figure}[ht]
\resizebox{\hsize}{!}{\includegraphics{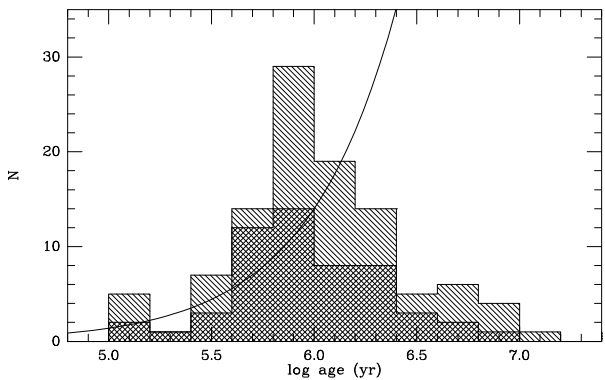}}
\caption{The age distribution of pre-main sequence stars near the Taurus molecular cloud.
Single-hatching denotes weak-emission T Tauri stars (WTTS), cross-hatching denotes 
strong-emission T Tauri stars (CTTS).  The curve indicates the expected distribution of
the number of stars per unit interval in log age for a constant star formation rate.
}
\end{figure}

The relatively narrow 1-2 Myr age spread in Taurus is difficult to understand for two
reasons.  First, if ambipolar diffusion is important, it is difficult to arrange for
the stars to form in a coordinated way over such a short period, since the
ambipolar diffusion timescale of 5-10 Myr should vary by factors of order unity among
cores of differing magnetic fluxes, shielding from the interstellar ultraviolet radiation
field, etc. (e.g., Myers \& Khersonsky 1995).  Second, it is not immediately obvious how
molecular gas can be brought to the point of forming stars in a coordinated way on a timescale
that is much shorter than the crossing or dynamical time for the entire region
($\gtrsim 20$~pc/2~$\kms \gtrsim 10$~Myr).

A few of the stars in Figure 2 do appear to have ages of 3-10 Myr.  
However, it is necessary to consider these objects in the appropriate context.  
The curve in Figure 2 corresponds to the distribution
of ages expected for a {\em constant} rate of star formation (see Kenyon \& Hartmann 1990).
This comparison shows that even if the ages of the older stars are not in error,
{\em they cannot represent a significant episode of star formation}.
This point is often overlooked in observational papers which tend to quote the extremes
of stellar age distributions rather than typical dispersion.
It is also important to emphasize
that {\em every source of error, including incorrect attribution of membership,
will increase, not decrease, the apparent age spread.}  Considering the likely observational errors 
(Kenyon \& Hartmann 1990; see also Gullbring et al. 1998)
the stellar age distribution in Figure 2 is remarkably narrow.

The first issue to address is whether the stellar ages determined in this way
have been systematically underestimated,
either due to observational errors or to errors in theoretical evolutionary tracks.
In general, the largest observational error probably arises from the inclusion of binary
stars; since contraction down Hayashi tracks results in a luminosity - time relation
of approximately $L \propto t^{-2/3}$ (e.g. Hartmann 1998), an maximum 
overestimate of the stellar luminosity by a factor of two would lead to an overestimate of
the age by a factor of three.  Some of the ``youngest'' objects are now known to be binary  
and thus older (e.g., Simon et al. 1995; see following paragraphs).  
However, it can hardly be the case that all the stars in Taurus are
equal-luminosity binaries, 
and so the failure to take multiplicity into account is unlikely to increase
the {\em median} age of the stellar population by more than a factor of $\sim 1.5$ (Kenyon \& Hartmann 1990).

A more plausible possibility is that the theoretical evolutionary tracks are systematically in error.
Neglecting deuterium fusion, low-mass pre-main sequence stars contract down Hayashi tracks, and
the stellar age is proportional to the Kelvin-Helmholtz timescale (Stahler 1988)
\begin{equation}
t_{KH}~=~{3 \over 7} {G M^2 \over RL} ~\propto~ {M^2 T_{ef\!f}^2 \over L^{3/2}}\,,
\end{equation}
where I have indicated the dependence on the observable parameters, luminosity $L$ and 
effective temperature $T_{ef\!f}$, as well as on the stellar mass $M$.  
(This is the age assuming contraction from infinite starting radius; if instead stars
are formed with finite radius at a ``birthline'', ages are {\em overestimated} from
conventional Hayashi tracks; Stahler 1988.) The dependence of
the age on $L$ and $T_{ef\!f}$ directly is rather weak, and except for the binary problem
discussed in the previous paragraph, unlikely to skew or bias ages to a significant
extent.  The determination of the stellar mass is, on the other hand, quite probably
subject to systematic errors.  However, to account for the difference between the ``observed''
Taurus population age and the ambipolar diffusion timescale, an increase
in ages by factors of 5 to 10 are required, corresponding to an increase in stellar masses by factors of 2-3.
It would not be surprising to find effects of this magnitude near the substellar limit,
but it seems much less likely that mass errors are this large for the typical $\sim 0.5 \msun$ T Tauri
stars which make up the bulk of the population in Figures 1 and 2. 

While this issue ultimately will be resolved by mass measurements of pre-main sequence binaries, 
already there are important constraints from the Keplerian rotation of 
circumstellar disks measured in $^{13}$CO (Guilloteau, Dutrey, \& Simon
1999; Guilloteau \& Dutrey 1998; Dutrey et al. 1998; Close et al. 1999).
It is worth looking at these results in detail to get some idea of the magnitude of the
errors.  Consider first GG Tau, which is a hierarchical quadruple system; the
brightest close pair Aa and Ab are situated within the circumbinary disk.  Guilloteau
et al. (1999) find that the combined mass of Aa and Ab is $1.28 \pm 0.07 \msun$ from
the Keplerian motion of the disk.   White et al. (1999) have carried out a careful 
spectroscopic and photometric study of this system, and find that Ab is only slightly
later in spectral type, and slightly lower in luminosity, than Aa, so to first order
we may assume that Aa has a mass of approximately $0.7 \msun$.  White et al. also
estimate the effective temperature of Aa, a K7 star, to be approximately 4000K,
and its luminosity as $0.84 \lsun$.  Then, using the approximation that the
Hayashi track age (neglecting birthline effects and D fusion) is 1/3 of the Kelvin-Helmholtz
timescale (Hartmann 1998), 
\begin{equation}
t(Hayashi)~\sim ~{1 \over 7} {G M^2 \over RL} ~\sim~ 1.5~{\rm Myr}\,.
\end{equation}
This age is identical to that derived by White et al. (1999) from adopting Baraffe
et al. (1998) evolutionary tracks.
The CMA tracks and assumptions used in constructing Figure 2 
result in a mass estimate of about $0.5 \msun$ and an age $\sim 0.8$~Myr.  
While this is very nearly a factor of two increase in the age, it does not come close to
the correction needed to make ambipolar diffusion timescales relevant.

Similar tests can be made for three other Taurus systems (all assuming a distance of 140 pc).
For GM Aur, Dutrey et al. (1998) find two possible solutions, reflecting the uncertainty
in $\sin i$; $0.84 \pm 0.05 \msun$ or $0.62 \pm 0.08 \msun$, while the CMA tracks and calibrations
used in Figure 2 produce a mass estimate of $0.52 \msun$ for this object.  This results in changing
the CMA track age of 0.9 Myr to 2.3-1.3 Myr, depending upon which mass estimate is used.
Guilloteau \& Dutrey (1998) estimate a dynamical mass of $0.47 \pm 0.06 \msun$
for DM Tau, similar to the mass of $0.43 \msun$ used in Figure 2; this changes the age
from 1.5 to 1.9 Myr.  Finally, the system mass of $1.2 \msun$ estimated by Close et al. (1998)
for UY Aur AB changes the age from the 0.3 Myr CMA track value to 0.6 Myr.
In summary, the current data suggest that the ages in Figure 2 might be systematically 
underestimated, but by at most a factor of 2, and so the inconsistency with typical ambipolar 
diffusion timescales remains.

There are two other (admittedly weaker) reasons for preferring a young age for the Taurus stars.
First, the ratio of T Tauri stars to Class I protostellar sources in Taurus is about
10:1 (Kenyon et al. 1990, 1994).  Assuming typical T Tauri ages of 1-2 Myr and steady 
recent star formation, this implies that the protostellar (free-fall) timescale
is about 0.1 - 0.2 Myr, in good agreement with the basic theory of
collapse from thermally-supported protostellar clouds (Shu 1977).  It is also
consistent with infall rates estimated from the analysis of spectral energy distributions
(Adams, Lada, \& Shu 1987; Kenyon, Calvet, \& Hartmann 1993). 
If instead the TTS are really 10 Myr old, protostellar collapse timescales become
an order of magnitude larger than predicted by theory, which is hard to understand;
additional supporting forces generally result in a denser initial state and faster, not slower, collapse.

Second, the spatial clumping of stars in Taurus is much easier to understand if the stellar ages
are closer to 1 Myr than 10 Myr.  Some of the stellar ``filaments'' shown in Figure 1 are narrower than 1 pc;
unless gravitational binding is important (which seems unlikely),
at a typical velocity dispersion of $0.5 \kms$ for 1 pc scales from Myers \& Goodman (1988)
one would estimate an age of $\lessim 2$~Myr.  Conversely, a 10 Myr age would require an
unreasonably small stellar velocity dispersion, 0.1 $\kms$ (half the sound speed), 
to avoid having the stars diffuse too far.

Thus, it is difficult to avoid the conclusion that the main phase of star formation in Taurus 
started 1-2 Myr ago over a spatial extent of 20 pc or more.

\section{Are We Missing Post-T Tauri Stars?}

As mentioned in the previous subsection, the ambipolar diffusion picture would imply minimum
age dispersions of several Myr as a result of differing initial conditions.
Even in the absence of magnetic flux reduction, it is surprising to find coordination
of star formation on timescales much shorter than the crossing or dynamical time of the complex.
The next question is whether the older stars have been missed; for example,
it would have been very difficult for past objective prism surveys to 
detect the weak H$\alpha$ emission of the older pre-main sequence stars -- the ``Post T Tauri'' stars, or PTTS.

The first systematic effort to find Taurus PTTS was that of Herbig, Vrba, \& Rydgren (1986), 
who objective prism plates to search for Ca II H and K emission, which decays
much more slowly with increasing age than H$\alpha$ emission.  Indeed,
Herbig et al. even detected stars which turned out to be (foreground) members 
of the Hyades cluster (age $\sim 600$~Myr).  However, the new members of Taurus found in this
way were not significantly older than the previously-known population.

More recently, Brice\~no et al. (1998) surveyed 30 square degrees of the Taurus region with
very deep H$\alpha$ objective prism plates.  Although their spectral resolution was poor, Brice\~no et al.
were able to detect very weak H$\alpha$ emission, down to equivalent widths $\sim 3$\AA 
(comparable to those observed in Pleiades M stars), by the simple expedient of 
selecting every M star with potential emission, and then following up with
high-resolution slit spectra of these objects.  Most of these potential
candidates exhibited no H$\alpha$ emission; the stars with
weak H$\alpha$ emission and Li absorption (see \S 4,5) 
were only marginally older than the previously-known population.

In an attempt to select stars by criteria completely independent of emission strength,
we (Hartmann et al. 1991) used a proper motion study conducted by Burt Jones to select candidates
for followup slit spectroscopy.  Although this study covered only a modest area of Taurus
(9 square degrees), it is significant that none of the new members identified in this way were
part of a significantly older population.

In contrast, X-ray surveys, first made with the Einstein satellite (Walter et al. 1988), and then
later with ROSAT, particularly the ROSAT All-Sky Survey (RASS; Neu\-h\"au\-ser et al. 1995)
did find a substantial number of older, X-ray active stars, many with substantial Li absorption
suggesting youth.  The main reason why these X-ray surveys ``succeeded'' 
where the optical objective prism and proper motion studies had previously ``failed''
is that the X-ray surveys cover much larger areas on the sky than the optical surveys.
For example, the original ``Taurus'' RASS survey of Neuh\"auser et al. (1995) covered 1000 square
degrees, vs. the 60 square degrees covered by Brice\~no et al. (1998) and the 10 square degrees
surveyed in the proper motion study.  Even the later followup study of ``Taurus'' by Wichmann et al. (1996)
surveyed a region of about 280 square degrees.  A second reason is that a number of the older
X-ray sources are relatively early, bright stars that tended to be overexposed on the optical plates.

Though this population discovered by X-ray surveys was suggested to be the answer to the missing
PTTS, the large volume of space encompassed by these stars compared to that occupied by the molecular gas
immediately calls into question whether the X-ray stars are really associated
with the present cloud complex.  To address this issue it is essential to
consider how WTTS and/or PTTS are defined.

\section{What is a WTTS/PTTS?}

In retrospect, the inadequacy of our present (spectroscopic) definition of a weak-emission 
T Tauri star (WTTS) has led to a great deal of confusion.
Originally, T Tauri stars were defined as having strong optical emission lines,
particularly H$\alpha$, Li I absorption, and spatial location near nebulosity or dark clouds.
We now know that there are weak-emission T Tauri which have ages similar to those
of the strong-emission or classical T Tauri stars (CTTS), but which have very weak H$\alpha$
emission (Figure 2), and may not lie in or near molecular gas and dust.  

It is clear from Figure 2 that the WTTS constitute an important segment of the pre-main sequence
population (roughly half of the stars with ages comparable to 1 Myr; Strom et al. 1989). 
Any definition of a T Tauri star must encompass these objects.
On the other hand, eliminating strong H$\alpha$ emission and position near 
molecular clouds as requirements, and relying on Li absorption and strong X-ray activity
causes problems because the definition now includes some {\em main sequence} stars.
Li is a very useful spectroscopic discriminant for stars with
masses $\lessim 0.6 \msun$; in these completely convective stars, Li is depleted by fusion
in the pre-main sequence phase
at ages $\sim 10$~Myr.  However, higher-mass stars do not deplete Li (much) in pre-main sequence
evolution, because the growth of the radiative core reduces the mixing of core and envelope material,
and thus reduces the envelope Li depletion.  For this reason
the presence of Li cannot be used very easily to distinguish pre-main sequence from main
sequence stars  (Brice\~no et al. 1997).  In fact, if the only spectroscopic requirement
is that Li be relatively undepleted, G and early K main-sequence stars in the Pleiades 
Soderblom et al. 1993) are {\em also} WTTS (or PTTS)!  

The G and early K stars in the Pleiades, with ages of $\sim 100$~Myr, also
have nearly identical X-ray properties to those of T Tauri stars, so there is no obvious way
of distinguishing 1 Myr-old WTTS from 100 Myr-old ``WTTS'' or ``PTTS'' from spectroscopic
or magnetic activity criteria alone.  One must have accurate distances and photometry to place
these stars in the HR diagram to derive an age; often the distances are not known.
It is not acceptable to {\em assume} that the X-ray stars are at the distance of Taurus if they
have similar radial velocities, because the velocity dispersions of stars with ages $\lessim 100$~Myr
are relatively small anyway.

The important question is {\em not} whether there are large numbers
of WTTS (or PTTS) as defined above; there must be many more such ``WTTS'' in the solar
neighborhood, because there must be many more 100 Myr old stars than 1 Myr old stars.
The true question of interest for understanding star formation
is whether there are substantial numbers of older stars 
{\em which originated in the present molecular cloud complex.}
Specifically, the ``missing'' 3-10 Myr-old stars must be found for 
Taurus to have had a continuous star forming history comparable to ambipolar diffusion and crossing times.

\section{Most RASS Sources are not from Taurus}

There is currently general agreement that many RASS sources are older than 10 Myr,
especially in directions not confused with star-forming regions. For example, Guillout et al.
(1998) summarize the RASS evidence for a significant population of stars in a ``Gould's Belt''
spatial distribution.  Guillout et al. draw no firm 
conclusion about whether this population of ages 30-80 Myr contaminates the ``Taurus''
samples; they argue that the Gould Belt in this direction may be too distant to contribute
significant numbers of stars.  Yet the general conclusion that low-mass stars of ages 
$\lessim 100$~Myr can contribute strongly to RASS populations, plus the possibility of
a Gould Belt distribution which passes {\em out} of the galactic plane through the
general position of the Taurus clouds on the sky, emphasizes the importance of careful
analysis of stellar populations before drawing sweeping conclusions.

The weight of the evidence clearly indicates that the distributed X-ray stars are not
mostly from the present-day Taurus molecular clouds, but instead constitute an older population
which must have originated in other molecular clouds.  The specific lines of argument are:
 
1. X-ray emission decays only slowly with age for G and early K stars younger than 100 Myr.  
This means that any X-ray survey covering
an appreciable volume in the solar neighborhood {\em must} include many stars older than 10 Myr
(unless there has been no star formation between 10 and 100 Myr ago, an implausible suggestion inconsistent
with, for example, the existence of nearby clusters like the Pleiades and $\alpha$ Per).
Brice\~no et al. (1997) used the Miller-Scalo (1979) average stellar birthrate
for the solar neighborhood, the detailed X-ray emission of stars as a function of age, and
the sensitivity of the RASS, and showed that the number of the stellar RASS sources in
Neuh\"auser et al. (1995) and Wichmann (1996) could be accounted for by star formation over a period
of 100 Myr.  This suggests that the correction that needs to be made to the RASS source
numbers to account for stars older than 10 Myr is of the same order as the total sample.
Favata et al. (1993) and Micela et al. (1997) made essentially the same point in their 
discussion of Einstein X-ray source statistics, and Favata et al. (1997) made this conclusion 
with respect to the RASS results as well.  (Similar findings have been made for Chamaeleon by Covino et al. (1997)).

2. The RASS survey shows clearly that the PTTS population is widely dispersed and fairly
uniformly distributed (e.g., Wichmann et al. 1996), with no evidence of clustering.
Now, any population of 3-10 Myr old stars should have dispersed no further than 6-20 pc from
the clouds (with a velocity dispersion of $2 \kms$), and this is roughly comparable to 
the extent of the present clouds.  In other words, a significant number of
the RASS sources should be (loosely) clustered near the existing clouds.  This is not observed.
Almost all of RASS sources which are found projected on or near molecular gas are very young,
i.e. they are WTTS similar to the 1 Myr WTTS population already known (Brice\~no et al. 1997).
The nearly uniform distribution of the RASS stars is much more consistent
with an older population originating in many different molecular clouds 
as suggested by Brice\~no et al. (1997); in this case, the relatively uniform spatial distribution
is due to formation in multiple sites, plus the spatial diffusion of stars over
30-100 Myr.

Some simple quantitative estimates help to illustrate the difficulty with interpreting the RASS sources
as the missing PTTS.
There are nearly 200 known young pre-main sequence stars in Taurus, most with ages less than 2 Myr and
concentrated within a region of about 20 pc in diameter.
Now, if Taurus has been forming stars at a more or less constant rate
for 10 Myr, there should be roughly 800 more stars with ages between 2 and 10 Myr;
with a velocity dispersion of only $2 \kms$, these stars could not have wandered 
far from the existing complex.  It is hard to see why the objective prism 
(60 square degrees) and proper motion surveys have found essentially no evidence for such a large population.
Even restricting the mass range to $\sim 1 - 0.8 \msun$, to include only the G and early K stars
that the RASS is most sensitive to among older stars (Brice\~no et al. 1997), and 
assuming a typical IMF, 
there should be $\sim 160$ stars in this mass range with ages between 2 and 10 Myr 
and within $\sim 20$~pc of the cloud complex.
Surveying an area of 280 square degrees with the RASS (corresponding to about $40 \times 40$~pc
at the distance of Taurus, or a much larger area),
Wichmann et al. (1996) only found about 76 new WTTS with Li, of these only 45 lie in the appropriate
range of spectral types ($\sim$ G-K4).  This number is considerably smaller than predicted, even assuming that
{\em none} of these stars are older than 10 Myr; but it is likely that many, if not most, of these stars are older,
for the reasons mentioned in item 1 (above) and item 3 (below).

3. Detailed estimates of ages from Li depletion, made using high-resolution spectra
and considering the mass dependence of depletion (Favata et al. 1997; Brice\~no
et al. 1997; Covino et al. 1997; \S 4), provide little evidence for a substantial PTT population.
The most detailed analysis is that of Martin \& Magazzu (1998), who studied 35 of the 76 Wichmann et al.
(1996) stars.  Of the 15 stars with spectral types of G0-K4 that Martin \& Magazzu studied, 
only 3 had Li abundances greater than that of stars in the Pleiades 
(with perhaps another one or two borderline cases).  Applying this
success rate to the entire Wichmann et al. sample suggests that they found 6-10
G-K4 or $\sim 1 - 0.8 \msun$ stars younger than the Pleiades.  Even if all these stars lie in the 3-10 Myr
age range (which is unlikely, since younger than the Pleiades only means
younger than $\sim 100$~Myr), it is obvious that they cannot represent an important period
of star formation.

At later spectral types,
Martin \& Magazzu (1998) recovered a higher fraction of stars with less Li depletion than in the Pleiades;
this is because X-ray emission decays with age faster in the lower mass stars, and so older stars in this
mass range drop out of the RASS (Brice\~no et al. 1997).  Of the remaining 20 stars in the sample with spectal
types K5 and later, 7 were WTTS (essentially no Li depletion)
and one was even a CTTS.  It is not yet clear whether these WTTS have ages in
the 3-10 Myr age range, or have the same ages as the previously-known population.
Martin \& Magazzu found a further 8 lower-mass stars with modest Li depletion, which they called ``PTTS''. 
It is unclear whether these objects have ages closer to the onset of Li depletion ($\sim 10$~Myr)
than to the age of the Pleiades. 

4. Radial velocities and proper motions generally do not clearly
distinguish stellar groups of ages $\lessim 100$~Myr, because the velocity dispersions
of such stars with respect to the local standard of rest are smaller ($\lessim 10 \kms$) 
than the solar motion ($\sim 20 \kms$).  Even so,
Frink et al. (1997) studied the proper motions of RASS sources found south of Taurus, and
concluded that they were incompatible with a Taurus origin.
There is kinematic evidence for an older population ($\sim 50$~Myr) of stars in
a ``Cas-Tau'' association (Walter \& Boyd 1991);
de Zeeuw et al. (1991) found that the Cas-Tau members have distinctly different proper motions
from those of pre-main sequence objects in Taurus.

In summary, while there may be a modest number of 3-10 Myr old stars formed near or in
Taurus, the vast majority of the Taurus population has been formed in the last 2 Myr or so,
and most of the RASS sources were formed in other star forming regions.

\section{Other solutions?}

Feigelson (1996) recognized that the wide dispersal of the RASS sources was essentially
impossible to explain with a 10 Myr old population, given the low velocity dispersions in
star-forming regions.  Feigelson also recognized that the observations of Taurus and other
sites imply not only a rapid burst of star formation, but a rapid
dispersal of star-forming gas as well; otherwise, not only would there be older stars, but there
would be a wider range of stellar ages than typically found.  Feigelson's solution to
this problem was to suggest that most of star formation occurs in small clouds which are
much more easily dispersed on short timescales.  But it is relatively easy to find CTTS 
projected on dark clouds; since most CTTS are not found in small, isolated dark clouds, 
this cannot be the explanation of the PTT problem.

In an attempt to save the ambipolar diffusion picture, Palla \& Galli (1997) suggested
that clouds are quiescent for 5-10 Myr before forming stars.  This explanation also
seems unlikely; as discussed above, there must be a range in initial conditions which results 
in a significant range of ambipolar diffusion timescales, making it difficult to coordinate
star formation on a much shorter timescale of 1-2 Myr.
This picture would also imply that there should be many more molecular clouds without than with
young stars, and this also seems not to be the case.

The simplest, and only viable explanation, of the Taurus observations 
is that star formation is rapid.  This result is consistent with results for other regions
on differing scales (Elmegreen 2000; Elmegreen, this volume); it appears that the timescale
for star formation is generally of the order of a crossing time (even less in the case
of Taurus).  Rapid star formation, without long ambipolar diffusion timescales, is also consistent
with lifetimes of cloud cores as inferred from statistical estimates (Lee \& Myers 1999; Jijina,
Myers, \& Adams 1999).  

Because Taurus is still forming stars, one cannot estimate the cloud
lifetime.  However, there is some evidence for dispersal already around the
L1551 cloud, and especially for the group near L1544 ($l \sim 178, b \sim -10$). 
More generally, none of the nearby star-forming regions have much older ages, implying
not only that star formation is rapid, but that cloud dispersal is fast as well.
The oldest of the nearby star-forming complexes seems to be
Cha I, with a typical stellar age of 3 Myr (Lawson et al. 1996). Even in this case,
there is controversy over the distance, and
ages would decrease significantly if Cha I is at the same distance as Cha II ($\sim$ 200 pc;
see Gauvin \& Strom 1992; Hughes \& Hartigan 1992). 

In their study of CO in distant stellar clusters, 
Leisawicz, Bash, \& Thaddeus (1989) found that despite
problems of large beam sizes and poor age determinations, molecular gas probably is
removed from clusters on a timescale of order 10 Myr.
Recent studies of nearby moving groups (TW Hya: Kastner et al. 1997, Webb et al. 1999;
$\eta$ Cha: Mamajek, Feigelson, \& Lawson 1999) also suggest that molecular gas
is dispersed at ages $\sim 10$ Myr.  The Orion OBIa association is also estimated to have
an age of about 12 Myr (de Zeeuw et al. 1999); this region, spanning a projected distance of about 30 pc
on the sky, is devoid of molecular gas.  
In short, all the evidence from the solar neighborhood points to both short timescales for both
star formation and molecular gas dispersal.

\section{Implications for star formation}

One implication of this picture of rapid star formation and cloud dispersal is that
magnetic fields are unlikely to play a dominant role and ambipolar diffusion is not
the primary process controlling the rate of star formation.  This is not to say that magnetic
fields do not have important dynamic effects, nor that ambipolar diffusion never occurs
or plays no role, but that other processes are more important.

Eliminating star formation in strongly magnetically-subcritical clouds actually reduces
problems for theory.  A subcritical cloud, by definition,
is one in which the restraining forces of the (static) magnetic field are stronger than gravity.
By force balance, this means that a subcritical cloud will expand, especially when internal
gas and turbulent pressures are included, {\em unless} the cloud is confined by {\em external}
pressures (Fiedler \& Mouschovias 1993).  This immediately raises two difficulties.
First, if the external confining medium has appreciable mass (as in the core-envelope
models of Mouschovias and collaborators), it is possible that 
while the cloud core is subcritical, the external envelope may be supercritical, in which case the entire configuration
may collapse and then fragment into smaller objects without ambipolar diffusion. 
Alternatively, a lower-density medium can confine the subcritical cloud due to turbulent
pressure.  But this mechanism also has severe problems, because numerical simulations
show that this turbulent external medium is almost certainly highly time-dependent,
with flows that can distort and disrupt the cloud as well as compress it (Bal\-le\-steros-Pa\-redes,
Vazquez-Semadeni, \& Scalo 1999).

Most observations of magnetic fields are consistent with the idea that the magnetic
energy density is roughly in equipartition with other energy densities in the ISM
(Crut\-cher 1999).  This implies that cloud cores are likely to be roughly
critical, not strongly subcritical.  Ciolek \& Basu (1999) have argued that faster
collapse of cloud cores, on larger scales, can be reconciled with ambipolar
diffusion models if cloud cores are much closer to being magnetically critical; 
this reduces the amount of magnetic flux that must be removed before 
free-fall collapse can proceed.  But in this case ambipolar diffusion is no longer
a major factor in setting the timescale for star formation anyway.

If cloud cores are, on the average, roughly critical, it must be that some clouds are
supercritical and others subcritical.  It then seems likely that the supercritical clouds
will collapse and form stars first.  If then clouds are dispersed rapidly, either by
the energy input from the forming stars or by the flows producing the clouds in the
first place (see below), there may not be enough time for the subcritical clouds to
collapse.  In this picture, the low efficiency of star formation on the galactic
scale noted by Zuckerman \& Evans (1974) is not the result of slowed collapse by
magnetic fields, but is purely the result of the generally low efficiency of converting
molecular cloud material into stars; and this low efficiency is in part the result
of the rapid dispersal of molecular gas.

\begin{figure*}[ht]
\resizebox{17cm}{!}{\includegraphics{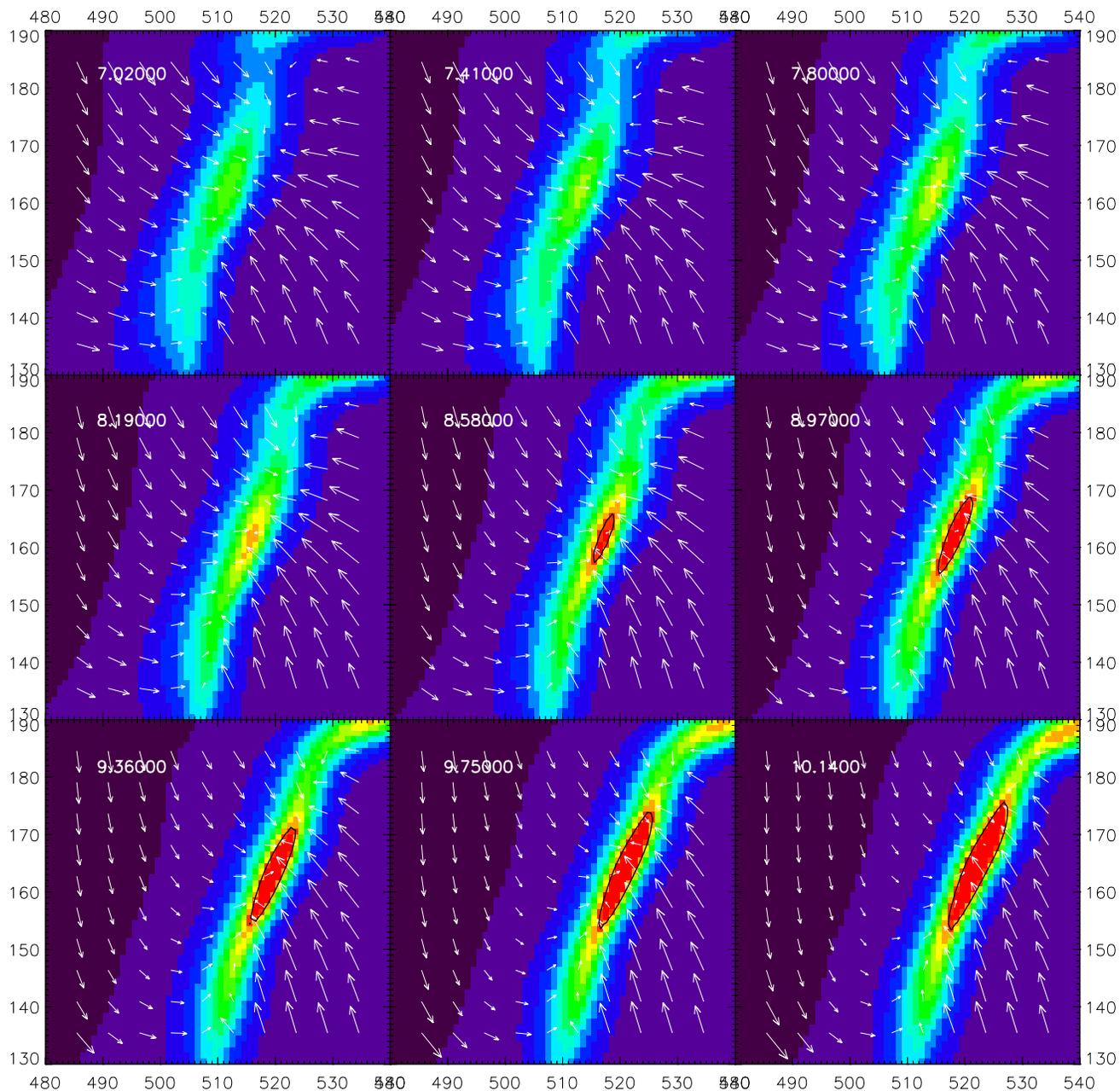}}
\vskip 6ex
\caption{Two-dimensional simulation of cloud formation from Ballesteros et al. (1999),
including magnetic fields.  The spatial scale is given in units of 1.26 pc, and the time
units in Myr are given in the upper left hand corner of each panel.
The solid contour marks the location of the highest-density regions $> 35 cm^{-3}$).
Converging flows of gas moving at $\sim 10 \kms$ collide to
form a cloud; the density increases rapidly once gravitational forces become important,
suggesting that molecular gas can be formed over a region of $\gtrsim 20$~pc on a timescale
of a few Myr.
}
\end{figure*}

\section{Rapid Cloud Formation and Dispersal}

As summarized by Elmegreen (2000) and in his contribution
elsewhere in this volume, observations are generally consistent
with molecular cloud lifetimes of order their crossing times.
The picture of rapid molecular cloud dispersal also eliminates the severe theoretical
problem of supporting molecular clouds by turbulence.  Recent numerical studies 
(Gammie \& Ostriker 1996; Stone, Ostriker, \& Gammie 1998; MacLow et al. 1998; MacLow 1999; 
Padoan \& Nordlund 1999) indicate that magnetic turbulence is rapidly damped,
and so that energy input has to be essentially continuous to support clouds by
this mechanism.  But the sources of energy input - stars, external flows - are likely
to be inhomogeneous and anisotropic in their effect.  Given a gaseous medium that
is effectively optically thin, and so cannot trap heat energy for long, it is difficult
to see how the required turbulent energy input can be ``fine-tuned'' to produce any
kind of equilibrium; one would expect either insufficient support, in which case collapse
ensues, or too much energy input, in which case the cloud is blown apart.  
It is much easier to assume that equilibrium generally does not exist,
and that cloud dispersal is sufficiently rapid that the issue of regenerating turbulent support 
never arises.

This does, however, raise a new difficulty; how can molecular clouds form so rapidly?
In particular, Taurus appears to pose special problems, because star formation appears
to have started in a coordinated way over a timescale of about 2 Myr or less across a region
with a dynamical time $\gtrsim 10$~Myr.  This seems 
explainable only with some kind of external ``triggering event''; but the Taurus clouds
are not obviously the remnant shell of an old supernova, or lie directly at the edge
of a spiral arm.  Suggestions have
been made about some kind of triggering as the cloud passed through the galactic plane its
present position, but the timescale for this mechanism is far too long.

In this context, the numerical 2D simulations of the galactic ISM by Passot,
Vazquez-Semadeni, Balle\-steros-Paredes, and other collaborators (Passot et al. 1995;
Balle\-steros-Paredes et al. 1999; Balle\-steros-Pa\-redes, this volume) are particularly interesting.
By simulating the large scale properties of the ISM within a 1 Kpc square shearing box, 
it is possible to consider the formation of molecular clouds
without excessive concern that the choice of boundary conditions 
completely dictates the results.  While these simulations have their limitations, 
the resulting ISM properties are reasonably consistent with observations.
These simulations indicate that the atomic gas
in the solar neighborhood is highly turbulent over a large range of size scales;
in particular, there are large-scale, organized {\em flows} of atomic material moving
at 5-10 $\kms$.  When these flows converge, clouds are built up.  In our recent paper
(Balle\-steros-Pa\-redes, \, Hartmann, \, \&\,  Vaz\-quez-Sema\-deni \, 1999) we showed that such flows
could in principle produce clouds which become dense enough to form stars on short
time\-scales of a few Myr (Figure 3).  In effect, this is a sta\-tistical or stochastic
version of the supernova triggering model; multiple sites of star formation put energy
into the ISM, and in certain places organized flows are built up which can be fairly coherent
over scales of many tens of pc - as required for Taurus.  

These simulations suggest that there need
be no specific previous star forming event which triggered formation of the Taurus clouds.
The problem of the short triggering time is reduced by the recognition that the important velocity
in the problem is {\em not} the (post-shock) velocity dispersion of the molecular gas,
but the 5-10 $\kms$ velocity of the atomic flows forming the molecular cloud.
This picture suggests that to understand the life cycle of molecular material near the solar
circle, it is necessary to trace the conversion of atomic gas into and from molecular gas in dynamic
processes.

It might be objected that the quiescence of molecular cloud cores and Taurus, and the agreement
of collapse timescales and Class I SEDs with the theory of thermally-supported clouds
(\S 2), does not support a dynamic picture.  I don't think this is necessarily the case,
because the star-forming flows pass through shock(s) which reduce the turbulent velocities (Figure 3).
The densest cores in Taurus may represent a ``pause'' or slowing of dynamical motions in post-shock
gas, without necessarily being precisely in hydrostatic equilibrium.  In any event, it would seem
easier to explain the presumed prolate structure of most cores (Myers et al. 1991; Ryden 1996)
with dynamic rather than hydrostatic models.

There is much more that needs to be understood and tested about this dynamic model for
molecular cloud formation (see contribution by Ballesteros-Paredes in this volume).  
Ballesteros et al. (1999) attempted to relate the dynamics
of the H I gas to the CO clouds; while the observations were consistent with the idea
that converging H I flows are producing the molecular gas, as predicted by the simulations, 
much more work needs to be done to test the picture.  
On the theoretical side, there may be problems in forming molecular gas rapidly enough.  
While many things remain to be understood, both the ages of young stellar populations
and recent numerical studies of turbulence in molecular clouds are pushing us in the
direction of understanding star formation as a dynamic process.

{\acknowledgements

I wish to acknowledge my collaborators on the age distributions of young stellar populations, 
especially Cesar Brice\~no and John Stauffer.
The primary spur to relating the stellar age distributions to properties of the interstellar
medium has come from my collaboration with Javier Ballesteros and Enrique Vazquez.
This work has been supported in part by NASA Grant NAG5-4282.
}

\end{document}